\documentstyle[12pt]{article}
\title{A distinguished set of modes in an accelerated frame of
reference}
\author{Llu\'{\i}s Bel 
\\
\it Laboratoire de Gravitation et Cosmologie Relativistes\\
\it CNRS/URA 769, Universit\'e Pierre et Marie Curie\\
\it 4, place Jussieu. Tour 22-12. Bo\^{\i}te courrier 142\\
\it 75252 PARIS Cedex 05, France}
\date{}
\begin{document}

\maketitle

\begin{abstract}
We construct a distinguished set of positive and negative energy modes
of the Klein-Gordon equation for any Born frame of
reference in Minkowski's space-time. Unlike the case of a galilean
frame of reference it is unclear whether this set of modes may be an appropriate basis to define the vacuum of a quantized scalar field in an accelerated cavity.
\end{abstract}

a.- {\it Klein-Gordon equation}. Given any space-time with line element:

\begin{equation}
\label {1.1}
ds^2=g_{\alpha\beta}(x^\rho)dx^\alpha dx^\beta, \quad 
\alpha,\beta,\cdots=0,1,2,3
\end{equation}
the Klein-Gordon equation for a classical field $\psi(x^\rho)$ reads,
using a system of units such that $c=1$:

\begin{equation}
\label {1.2}
(\Box-\frac{m^2}{\hbar^2})\psi=(g^{\alpha\beta}\partial_{\alpha\beta}
-\Gamma^\alpha\partial_\alpha-\frac{m^2}{\hbar^2})\psi=0
\end{equation}
where:

\begin{equation}
\label {1.3}
\Gamma^\alpha=g^{\lambda\mu}\Gamma^\alpha_{\lambda\mu}
\end{equation}	
If $\psi_1$ and $\psi_2$ are two, in general complex, solutions then
the current:

\begin{equation}
\label {1.4}
J^\alpha(\psi_1, \psi_2)=
-i\hbar g^{\alpha\beta}(\psi^*_1\partial_\beta\psi_2
-\psi_2\partial_\beta\psi^*_1)
\end{equation}
is conserved:

\begin{equation}
\label {1.5}
\nabla_\alpha J^\alpha
=\frac{1}{\sqrt{-g}}\partial_\alpha(\sqrt{-g}J^\alpha)=0, \quad 
g=det(g_{\alpha\beta})
\end{equation}
and this allows to define the invariant scalar
product $(\psi_1,\psi_2)$ of two well behaved solutions as the flux
of the preceding current across any space-like hypersurface 
${\cal S}(t_0):t=t_0$

\begin{equation}
\label {1.6}
(\psi_1,\psi_2)=\int_{{\cal S}(t_0)}J^\alpha d{\cal S}_\alpha (t_0)
\end{equation}

b.- {\it Quantization of a scalar field}. The canonical quantization
of a scalar field is a two step process. The first step consists in
selecting a distinguished set of modes of the Klein-Gordon equation
to define what in the jargon of Quantum field theory is called the
{\it Vacuum state}. The second step consists in implementing the
so-called canonical commutation relations to be satisfied by the
field operator and its conjugate momentum. This paper is entirely
dedicated to the first step. A common belief\,\footnote{See for example
\cite{BD}, \cite{Fulling}} is that in general 
there is no unique way of choosing a unique vacuum state
and therefore that there is an inevitable spontaneous creation of
particles. In a preceding paper we claimed that for Robertson-Walker models
with flat space-sections it is possible to distinguish a preferred
set of modes, with no time mixing of positive and negative modes,
thus defining a unique vacuum state and suppressing the spontaneous
particle creation out from the vacuum state. In this paper we consider
the case of any Born motion in Minkowski's
space-time.

In Minkowski space-time and in a galilean frame of reference the
vacuum state is defined by the set of modes $(\epsilon=\pm)$:

\begin{equation}
\label {1.31}
\varphi_\epsilon(x^\alpha, \vec k)=(2\pi\hbar)^{-3/2}
u_\epsilon(t,\vec k)e^{\frac{i}{\hbar}\vec k \vec x}, \quad
\varphi_{-\epsilon}(x^\alpha, \vec k)=\varphi_\epsilon^*(x^\alpha, -\vec k)
\end{equation}	
with:

\begin{equation}
\label {1.32}
u_+(t, \vec k)= (2\omega)^{-1/2}e^{-\frac{i}{\hbar}\omega t} \quad 
\omega(\vec k)=+({\vec k}^2+m^2)^{1/2}
\end{equation} 		  
where $\vec k$ is a constant index vector. The modes $\varphi_+$ are
by definition the positive energy modes and $\varphi_-$ the negative
energy modes. This set of modes can be characterized by the following 
conditions:

i) The set of modes, having the general form \ref{1.31}, must be a
complete orthonormal set of particular solutions of the Klein-Gordon
equation. Orthonormality here means that the scalar product of two modes
is:

\begin{equation}
\label {1.33}
(\varphi_{\epsilon_1}(x^\alpha, {\vec k}_1),
\varphi_{\epsilon_2}(x^\alpha, {\vec k}_2))=
\frac{1}{2}(\epsilon_1+\epsilon_2)\delta_{\epsilon_1\epsilon_2}
\delta({\vec k}_1-{\vec k}_2)
\end{equation}
and completeness means that any solution of the Klein-Gordon equation
that can be written as a Fourier transform on the flat space-sections
$t=const$:

\begin{equation}
\label {1.34}
\psi(t, \vec x)=\frac{1}{(2\pi\hbar)^{3/2}}\int	
c(t, \vec k)e^{\frac{i}{\hbar}\vec k \vec x} \, d^3\vec k
\end{equation} 
can also be written as:

\begin{equation}
\label {1.35}
\psi(t, \vec x)=\int	
(a_+(\vec k)\varphi_+(x^\alpha, \vec k)
+a_-(\vec k)\varphi_-(x^\alpha,\vec k)) \, d^3\vec k
\end{equation}

ii) The functions $u_\epsilon$ are solutions of the first order
differential equation:

\begin{equation}
\label {1.36}
i\hbar\dot u_\epsilon=\epsilon\omega u_\epsilon, \quad \dot u=\frac{du}{dt} 
\end{equation}
It is this condition that guarantees that there will not be time
mixing of positive and negative modes. Generalizing this second
condition to a Born frame reference is the main contribution of this
paper. 

c.-{\it Born frames of reference}. Let $\cal C$ be a time-like
congruence of Minkowski space-time ${\cal M}_4$ and let $u^\alpha(x^\rho)$ be the
unit tangent vector field to $\cal C$. Born congruences are those
congruences satisfying the conditions,\cite{Born}:

\begin{equation}
\label {2.1}
{\cal L}(u^\rho)(g_{\alpha\beta}+u_\alpha u_\beta)=0
\end{equation} 
where ${\cal L}(u^\rho)$ is the Lie derivative operator along $u^\rho$
and $g_{\alpha\beta}$ is the space-time metric of ${\cal M}_4$. All
these congruences are either Killing congruences or irrotational,
\cite{Herglotz}, \cite{Noether}.

Let $\cal P$ be a domain of ${\cal M}_4$ where the congruence $\cal C$ is well defined, i.e. there exists only one world-line passing through any event of $\cal P$. We shall say that $\cal P$ is a proper domain if $\cal P$ contains in its entirety each of its world-lines, supposed to have infinite proper length and finite intrinsic curvature. We shall say that a proper domain $\cal G$ is a global domain if there is no other Born congruence ${\cal C}^\prime$ which contains $\cal C$ and its proper proper domain ${\cal P}^\prime$ contains $\cal G$.

Considering any world-line $\cal W$ of a Born congruence and using Fermi
coordinates based on $\cal W$ the line element of ${\cal M}_4$ is:

\begin{equation}
\label {2.2}
ds^2 = -[1+a_i(t)x^i]^2dt^2 + \delta_{ij}dx^i dx^j
\end{equation}
where $t$ is proper time along $\cal W$ and $a_i(t)$ is its intrinsic
curvature. The coordinate transformations:

\begin{equation}
\label {2.3}
x^{\prime i}=x^i-\lambda ^i\qquad 
t^{^{\prime }}=t +v_k(t)\lambda ^k 
\end{equation}
where: 
\begin{equation}
v_k(t)=\int_0^t a_k(u)\,du 
\end{equation}
leave invariant the compound description of the line-element
\ref{2.2} as a function of the coordinates $x^i$ and the function
$a_i(t)$ \,\footnote{We defined the concept of compound description
and the related concept of generalized isometries in \cite{Salas}} . 
In other words:

\begin{equation}
\label {3.1}
ds^2 = -[1+a^\prime_i(t^\prime)x^{\prime i}]^2dt^2 
+ \delta_{ij}dx^{\prime i} dx^{\prime j}
\end{equation} 
with:

\begin{equation}
\label {3.2}
a^\prime_i(t^\prime)
=\frac{a_i(t^\prime)}{1+a_k(t^\prime)\lambda^k}
\end{equation}
The old coordinates $(t, x^i)$ are the Fermi coordinates based on the
world-line $x^i=0$ and the new coordinates $(t^\prime, x^{\prime i})$
are the Fermi coordinates based on the world-line $x^i=\lambda^i$,
i.e. $x^{\prime i}=0$.

Using any of the Fermi coordinates defined above the Klein-Gordon 
equation in a Born Frame of reference can be written as:

\begin{equation}
\label {2.4}
\hbar^2(-\partial^2_t+A\partial_t +\zeta^{-2}\triangle+B^i\partial_i)\psi 
-\zeta^{-2}m^2\psi=0
\end{equation} 
where:

\begin{equation}
\label {2.5}
\zeta=(1+a_sx^s)^{-1}, \quad A=\zeta\dot a_j x^j, \quad
B^i=\zeta^{-1}a^i
\end{equation}
and:

\begin{equation}
\label {4.1}
\triangle=\delta^{ij}\partial_{ij}, \quad a^i = a_i, \quad 
\dot a =\frac{da}{dt}
\end{equation}

Let ${\cal S}(t_0)$ be the euclidean space-section $t=t_0$ and let $\cal G$ be the a global domain of the congruence. We shall write 
${\cal D}(t_0)={\cal G}\cap{\cal S}(t_0)$ and define the scalar product of two solutions of \ref{2.4} defined on $\cal G$ as:

\begin{equation}
\label {2.6}
(\psi_1,\psi_2)=i\hbar
\int_{\cal D}\zeta(\psi^*_1\partial_t\psi_2-\psi_2\partial_t\psi^*_1)\, d^3x
\end{equation}
If the two solutions are such that $\zeta^{-1} J^i(\psi_1,\psi_2)$ is zero on the frontier of $\cal G$ then the scalar product thus defined is independent of $t_0$.

e.-{\it Time reduction of the Klein-Gordon equation}.
Let $\hat F$ be in general a time dependent linear operator. We
shall say that the operator is an space operator if it commutes with
the product of a function which does not depend on the space
coordinates $x^i$:

\begin{equation}
\label {3.4}
\hat Fc(t)=c(t)\hat F
\end{equation}
We shall define the indicial function $F(t,x^i,k_j)$ of an 
space operator by the following formula:

\begin{equation}
\label {3.5}
\hat F\exp(\frac{i}{\hbar}k_jx^j)=
 F(t,x^i,k_j)\exp(\frac{i}{\hbar}k_jx^j)
\end{equation}
If $\hat F$ and $\hat G$ are two space operators then we shall
define the deformed product $( F\circ  G)(t,x^i,k_j)$ of the 
respective indicial functions as the indicial function of the product of the 
two operators, i.e.:

\begin{equation}
\label {3.6}
\hat F\hat G\exp(\frac{i}{\hbar}k_jx^j)= 
( F\circ  G)\exp(\frac{i}{\hbar}k_jx^j)
\end{equation}
The deformed product thus defined is associative. The general formula
to calculate it is:

\begin{equation}
\label {3.7}
 F\circ  G=\frac{1}{(2\pi\hbar)^3}\int\int
 F(t,x^r,k^\prime_s) G(t,x^r+\lambda^r,k_s)
\exp(\frac{i}{\hbar}((k_j-k^\prime_j)\lambda^j)
\,d^3k^\prime \,d^3\lambda
\end{equation}
This formula can be expanded in power series of $i\hbar$ (Hint: Use the variables of integration $\mu^j=\lambda^j/\hbar$ and develope the integrand of the remaining equation in powers of $\hbar$). The result is:

\begin{equation}
\label {3.8}
 F\circ  G=  F G+\cdots+
\frac{(-i\hbar)^n}{n!}
\frac{\partial^n  F}{\partial k_{s_1}\cdots\partial k_{s_n}}
\frac{\partial^n  G}{\partial x^{s_1}\cdots\partial x^{s_n}}+\cdots
\end{equation}
It is the property

\begin{equation}
\label {3.9}
\lim_{\hbar \to 0} F\circ  G
= F G
\end{equation}
that justifies calling $\circ$ a deformed product and $\hbar$ the
parameter of the deformation \,\footnote{We used a similar deformed product in another context in Ref. \cite{DefProduct}}.

The equation \ref{2.4} can be written as:

\begin{equation}
\label {2.7}
{\hat L}^2\psi =\hat W\psi
\end{equation}
where $\hat L$ and $\hat W$ are the linear differential operators:

\begin{equation}
\label {2.8}
\hat L=i\hbar(\partial_t+E), \quad 
\hat W=\hbar^2(-\zeta^{-2}\triangle -B^i\partial_i
-C)+\zeta^{-2}m^2
\end{equation}
with $A$ and $B^i$ given by \ref{2.5}, and:

\begin{equation}
\label {3.3}
E=-\frac{1}{2}A, \quad C=\partial_tE+E^2
\end{equation}

Since $\hat W$ does not contain any time derivative operator
it is an space operator. Its indicial function is:

\begin{equation}
\label {3.10}
 W=\zeta^{-2}({\vec k}^2 +m^2)-i\hbar B^jk_j
-\hbar^2(\partial_t E+E). 
\end{equation}
$\hat L$ is the sum of a time evolution operator $i\hbar\partial_t$ and 
$i\hbar E$
which is an ordinary multiplication operator. The
latter is therefore also an space operator.

Writing the Klein-Gordon equation as in \ref{2.7} suggests 
the possibility of constructing a first order time reduction of it, i.e,
an equation of the following type:

\begin{equation}
\label {2.9}
\hat L\psi = \hat f\psi
\end{equation}
$\hat f$ being a linear space operator
such  that each
solution of the equation above be a solution of \ref{2.4}. Notice that Eq.
\ref{2.9} is indeed a first order equation with respect to the
time coordinate if $\hat f$ is an space operator because in this
case the r-h-s of \ref{2.9} has a meaning for each value of $t=t_0$ once 
$\psi(t_0, x^i)$ is known. 

Acting on both sides of \ref{2.9} with $\hat L$ we obtain:

\begin{equation}
\label {2.10}
{\hat L}^2\psi= \hat L\hat f\psi =([\hat L, \hat f] + \hat f \hat L)\psi
\end{equation}
where $[\hat L, \hat f]$ is the commutator of the two operators.
Therefore using again eqs. \ref{2.9}, and \ref{2.7} we see that the
first will be a first order time reduction of the second one {\it iff} we
have:

\begin{equation}
\label {2.11}
([\hat L, \hat f] + {\hat f}^2)\psi= \hat W\psi
\end{equation}
for every solution of \ref{2.9}. 

We are interested in functions $\psi(t,x^i)$ that have a
Fourier decomposition:

\begin{equation}
\label {2.15}
\psi(t,x^i)=\frac{1}{(2\pi\hbar)^{3/2}}\int
c(t,k_j)\exp(\frac{i}{\hbar}k_jx^j)\,d^3k; 
\end{equation}
then using the definitions above we have, $f$ being the indicial
function of the operator $\hat f$:

$$
\hat L\hat f\psi(x^\alpha)=\frac{i\hbar}{(2\pi\hbar)^{3/2}}\int
\left(\dot c(t,k_j) f(x^\alpha,k_s)
+ c(t,k_j)(\partial_t f(x^\alpha,k_s)+E(x^\alpha)\circ f(x^\alpha,k_s))\right)\cdot
$$

\begin{equation}
\label {3.11}
\cdot\exp(\frac{i}{\hbar}k_jx^j)\,d^3k
\end{equation}	
and:

\begin{equation}
\label {3.12}
\hat f\hat L\psi=\frac{i\hbar}{(2\pi\hbar)^{3/2}}\int
\left(\dot c(t,k_j) f(t)+ c(t,k_j) f(t)\circ
 E\right)\exp(\frac{i}{\hbar}k_jx^j)\,d^3k
\end{equation}
We see from these expressions that the derivatives $\dot c(t, k_j)$ cancel 
out from the $[\hat L, \hat f]\psi$ and therefore the three operators
appearing in \ref{2.11}, $[\hat L, \hat f]$ $\hat f$ and $\hat W$ are space
operators. It follows from this that \ref{2.11} can hold for every
solution of \ref{2.9} {\it iff} it holds for every exponential function
$\exp(\frac{i}{\hbar}k_jx^j)$. This means that eq. \ref{2.11} is
equivalent to the following equation for the indicial function
$ f$

\begin{equation}
\label {2.12}
i\hbar(\partial_t  f+E\circ f- f\circ E)+ f\circ f= W
\end{equation}

Let us consider the following transformation:

\begin{equation}
\label {4.2}
 F^\#(t,x^i,k_j)= F^*(t,x^i,-k_j)
\end{equation}	
where $*$ means complex conjugation. A short discussion based on
\ref{3.7} shows that:

\begin{equation}
\label {4.3}
( F\circ  G)^\#= F^\#\circ  G^\#
\end{equation}
Using this property in eq. \ref{2.12} we get:

\begin{equation}
\label {4.4}
-i\hbar(\partial_t  f^\#+E\circ f^\#- f^\#\circ
E)+ f^\#\circ f^\#= W^\#
\end{equation}
But from \ref{3.10} it follows that:

\begin{equation}
\label {4.5}
 W^\#= W
\end{equation}
and as a consequence of that and from \ref{4.4} we can conclude that
if $f_+$ is a solution of \ref{2.12} then:

\begin{equation}
\label {4.14}
 f_-=- f_+^\#
\end{equation}
is again a solution of the same equation.

Let:

\begin{equation}
\label {4.6}
\psi_+(t,x^i)=\frac{1}{(2\pi\hbar)^{3/2}}\int
c_+(t,k_j)\exp(\frac{i}{\hbar}k_jx^j)\,d^3k
\end{equation}
be a solution of equation:

\begin{equation}
\label {4.7}
\hat L\psi_+ =\hat f_+\psi_+ 
\end{equation}
where $\hat f_+$ is the linear space operator whose indicial function is
some solution of eq. \ref{2.12}. The l-h-s is:

\begin{equation}
\label {4.8}
\hat L\psi_+=\frac{i\hbar}{(2\pi\hbar)^{3/2}}\int
(\dot c_+(t,k_j)+Ec_+(t,k_j))\exp(\frac{i}{\hbar}k_jx^j)\,d^3k 
\end{equation}
and the r-h-s is:

\begin{equation}
\label {4.11}
\hat f_+\psi_+=
\frac{1}{(2\pi\hbar)^{3/2}}\int
c_+(t,k_j)f_+(t,x^i,k_j)\exp(\frac{i}{\hbar}k_jx^j)\,d^3k
\end{equation}
From \ref{4.6} it follows that:

\begin{equation}
\label {4.9}
\psi_+^*(t,x^i)=\frac{1}{(2\pi\hbar)^{3/2}}\int
c_-(t,k_j)\exp(\frac{i}{\hbar}k_jx^j)\,d^3k
\end{equation}
where:

\begin{equation}
\label {4.10}
c_-(t,k_j)=c^*_+(t,-k_j)
\end{equation}
From this relation and from \ref{4.8} we have:

\begin{equation}
\label {4.12}
(\hat L\psi_+)^*=\frac{-i\hbar}{(2\pi\hbar)^{3/2}}\int
(\dot c_-(t,k_j)+Ec_-(t,k_j))\exp(\frac{i}{\hbar}k_jx^j)\,d^3k 
\end{equation} 
and from \ref{4.10} and from \ref{4.11}:

\begin{equation}
\label {4.13}
(\hat f_+\psi_+)^*=
\frac{-1}{(2\pi\hbar)^{3/2}}\int
c_-(t,k_j)f_-(t,x^i,k_j)\exp(\frac{i}{\hbar}k_jx^j)\,d^3k
\end{equation}
with $f_-$ defined as in \ref{4.14}. Equating the r-h-s of \ref{4.12}
to that of \ref{4.13} we prove that:

\begin{equation}
\label {4.15}
\hat L\psi_-=\hat f_-\psi_-
\end{equation} 
with:

\begin{equation}
\label {4.16}
\psi_-=\psi^*_+
\end{equation}
and $\hat f_-$ being the space operator with indicial function
$ f_-$.

Equation \ref{2.12} can be solved order by order assuming that
solutions exist that have a Laurent expansion in powers of $i\hbar$:

\begin{equation}
\label {4.17}
f=\sum_{n=-s}^{\infty}(i\hbar)^n f_n, \quad s<\infty
\end{equation}
Substituting this expression into \ref{2.12} we obtain to begin with:

\begin{equation}
\label {4.18}
f_n=0 \quad \hbox{for}\quad -s\le n<0
\end{equation} 
and:

\begin{equation}
\label {4.19}
f_0^2= W_0, \quad  W_0=\zeta^{-2}(k^2+m^2),\ \ 
k^2 =\delta^{ij}k_ik_j
\end{equation}
It follows that our assumption distinguishes two
particular solutions of eq. \ref{2.12}. Namely those solutions whose
starting terms in their development \ref{4.17} are:

\begin{equation}
\label {4.20}
f^+_0=+(W_0)^{1/2} \ \ \hbox{and} \ \ f^-_0=-(W_0)^{1/2}
\end{equation}
An examination of \ref{2.12} and \ref{3.8} shows that $f_n$ can be
calculated once $f_0,\cdots,f_{n-1}$ are known. Thus, for instance,
the next term $f_1$ is easily found to be:

\begin{equation}
\label {4.23}
f_1=-\frac{1}{2}\zeta{\dot a}_jx^j
\end{equation}
The two
expressions $f^{\pm}_0$ can be considered as the generators of two
solutions of \ref{2.12} that we shall note $f_{\pm}$. Since we
obviously have:

\begin{equation}
\label {4.21}
f^-_0=-(f^+_0)^*
\end{equation}
from the result stated in \ref{4.14} we conclude that:

\begin{equation}
\label {4.22}
f_-=-f_+^{\#}
\end{equation}
where now $f_+$ and $f_-$ are the two solutions generated by
\ref{4.20}. Equivalently we can write for any function $\psi$:

\begin{equation}
\label {6.1}
\hat f_-\psi^*=-(\hat f_+\psi)^*
\end{equation}

Let $\chi(t,x^i)$ be for
each value of t an eigen-state of the operator $\hat f_+$
corresponding to the eigenvalue $\rho(t)$:

\begin{equation}
\label {6.2}
\hat f_+\chi=\rho\chi
\end{equation}
From \ref{6.1} it follows immediately that $\chi^*$ is then an
eigen-state of the operator $\hat f_-$
corresponding to the eigenvalue $-\rho^*(t)$:

\begin{equation}
\label {6.3}
\hat f_-\chi^*=-\rho^*\chi^*
\end{equation}

f.-{\it Positive and negative energy modes}. 
Let us consider a space section ${\cal S}(t_0)$. We shall define the positive
(resp. negative) energy modes relatives to this section,
$\varphi_\epsilon(t_0;x^\alpha,k_j)$ with $\epsilon=+$ (resp.
$\epsilon=-$), as the solutions of the first order differential
equation:

\begin{equation}
\label {4.24}
\hat L\varphi_\epsilon=\hat f_\epsilon\varphi_\epsilon
\end{equation}
corresponding to the initial condition on $S$:

\begin{equation}
\label {4.25}
\varphi_\epsilon(t_0;t_0,x^i,k_j)=\frac{1}{(2\pi\hbar)^{3/2}}
\exp(\frac{i}{\hbar}k_jx^j)
\end{equation}

Let $\varphi_{\epsilon_1}$ and $\varphi_{\epsilon_2}$ be two modes, positives or negatives depending on the values of $\epsilon_1$ and $\epsilon_2$. The scalar product of these two modes calculated on the domain ${\cal D}={\cal G}\cap{\cal S}(t_0)$ defined in section c.- is:

$$
(\varphi_{\epsilon_1}(k_1),\varphi_{\epsilon_2}(k_2))=\frac{1}{(2\pi\hbar)^3}
\int_{\cal D}\zeta(x^\alpha)(f_{\epsilon_2}(x^\alpha,k_2)
+f^*_{\epsilon_1}(x^\alpha,k_1))\cdot
$$
\begin{equation}
\label{A1}
\cdot\exp(\frac{i}{\hbar}(k_{2i}-k_{1i})x^i)\,d^3x
\end{equation}
In general it is not possible to anticipate which pairs of modes will be orthogonal except in one case. Namely when $\epsilon_1=-\epsilon_2$ and $k_1=-k_2$:

\begin{equation}
\label{A2}
(\varphi_{+}(k),\varphi_{-}(-k))=0
\end{equation}
In this case in fact it follows from \ref{4.22} that the second factor in the integrand above will be zero.

The case where the Born congruence is such that the acceleration $a^i$ of each of the world lines does not depend on time is particularly simple\,\footnote{From a more conventional point of view, this case was considered in detail in \cite{Fulling2}}. In this case the exact solutions of \ref{2.12} are:
 
\begin{equation}
\label{A3}
f_\epsilon=\epsilon \zeta^{-1}(k^2+m^2)^{1/2}
\end{equation}
and therefore \ref{A1} becomes:

$$
(\varphi_{\epsilon_1}(k_1),\varphi_{\epsilon_2}(k_2))=\frac{1}{(2\pi\hbar)^3}
\int_{\cal D}(\epsilon_2(k_2^2+m^2)^{1/2}+\epsilon_1(k_1^2+m^2)^{1/2})\cdot
$$

\begin{equation}
\label{A4}
\cdot\exp(\frac{i}{\hbar}(k_{2i}-k_{1i})x^i)\,d^3x
\end{equation}
The integrand is the same that one has for a galilean frame of reference but the domain of integration is not. It follows from this that the orthogonality conditions \ref{1.33} will not be satisfied in this case either.  

From the result \ref{4.16} applied to $\varphi_\epsilon(t_0;x^\alpha, k_j)$ 
we get:

\begin{equation}
\label {4.27}
\hat L(\varphi^\#_+-\varphi_-)=
\hat f_-(\varphi^\#_+-\varphi_-)
\end{equation}
This is a first order differential equation for a variable which
satisfies on $S$:

\begin{equation}
\label {4.28}
\varphi^\#_+(t_0;t_0,x^i, k_j)-\varphi_-(t_0;t_0,x^i, k_j)=0
\end{equation}
We conclude then that:

\begin{equation}
\label {4.29}
\varphi^\#_+=\varphi_- \ \ \hbox{or} \ \   
\varphi^*_+(t_0;x^\alpha, -k_j)=\varphi_-(t_0;x^\alpha, k_j)
\end{equation}
recovering a property which the modes satisfy in a galilean frame of
reference.

Two questions have to be addressed now. The first one is the
following:
to write the line-element as in \ref{2.2} we used Fermi
coordinates based on some unspecified world-line of the frame of
reference. As indicated in the corresponding section one can switch
from a world-line to another with a coordinate transformation of the
type \ref{2.3}. Such transformations leave everything form-invariant
but the acceleration functions $a_i$ have to be modified according to
\ref{3.2}. The question then can be raised as whether the distinction
between positive and negative modes could depend on the Fermi
coordinates being used. From \ref{4.19}, \ref{4.20} and from \ref{2.5}
we see
that modifying the accelerations according to \ref{3.2} does not mix
the positive and negative determination of the generating first order
terms $f^\pm_0$. This proves that the distinction of positive and
negative modes is independent of the world-line on which the Fermi
coordinates are based.

The second question is whether the distinction between positive and
negative modes depend on the section on which one selects the
conditions \ref{4.25}. Let us consider a positive energy mode
$\varphi_+(t_1;t,x^i,k_j)$ satisfying the initial condition:

\begin{equation}
\label {5.1}
\varphi_+(t_1;t_1,x^i,k_j)=\frac{1}{(2\pi\hbar)^{3/2}}\exp(\frac{i}{\hbar}k_jx^j)
\end{equation} 
on some space-section $t=t_1$. On $t=t_0$ we shall have in general:

\begin{equation}
\label {5.2}
\varphi_+(t_1;t_0,x^i,k_j)=\int
a_+(t_1,t_0,k_j,k^\prime_s)\exp(\frac{i}{\hbar}k^\prime_jx^j)\,
d^3k^\prime
\end{equation}
Let us consider the following expression:

\begin{equation}
\label {5.3}
Z=\varphi_+(t_1;x^\alpha,k_j)-\int
a_+(t_1,t,k_j,k^\prime_s)\varphi_+(t_0;x^\alpha,k_j)\,
d^3k^\prime
\end{equation}
This auxiliary quantity satisfies the first order equation:

\begin{equation}
\label {5.4}
\hat L J=\hat f_+J
\end{equation}
and it is zero by construction on $t=t_0$. Therefore it will be zero
everywhere, meaning that a positive energy mode relative to any
space-section $t=t_1$ is always a linear superposition of a set of
positive modes relative to any other space-section. {\it Mutatis
mutandis} the same is true of course considering negative energy
modes instead of positive ones. This proves that although the
individual modes that we have defined depend on a space-section the
sub-spaces that the positive and negative modes span are
intrinsically distinguished. 

g.- {\it Conclusion}.
This paper has followed a path similar to that that we developed in a
preceding paper, \cite{Cosmology}, for the much simpler case of the
Robertson-Walker metrics. Here we have remained at a pure formal
level and whether the construction of the distinguished modes can be
achieved has not been neither proved, nor disproved of course.
One thing is unavoidable, namely that, except in the case of eq. \ref{A2}, the 
positive energy modes will not be orthogonal to the negative energy ones in
the sense of the natural scalar product. This would mean that either
this system is unsuited to be used in the quantization of scalar fields or
that the
canonical commutation relations between the scalar field and its
conjugate momentum should be generalized. Despite of all that we believe that the template that we have
presented can be useful as a starting point to anybody wishing to make his/her mind
about the problem of spontaneous particle creation in accelerated frames of
reference, or in some other cases of interest like for instance time dependent spherically symmetric gravitational fields.        

\section*{Acknowledgements}

It is a pleasure to acknowledge many stimulating and useful discussions with J. Llosa, A. Molina and P. Teyssandier.

\end{document}